\documentclass[journal]{IEEEtran}
\usepackage{graphicx}  
\usepackage{url}      
\usepackage{amsmath}                           
\usepackage{array}

\begin{document}
\title{Thickness Effects in the Resonance of Metasurfaces made of SRRs ans C-SRRs}%
\author{L.M. Pulido-Mancera$^{1}$, J.D. Baena$^{1}$, J.L. Araque$^{1}$\\1 Group of Applied Physics, Physics Department, Universidad Nacional de Colombia\\2 Electrical Engineering Department, Universidad Nacional de Colombia \\lmpulidom@unal.edu.co}
\markboth{2013 IEEE International Symposium on Antennas and Propagation and USNC-URSI National Radio Science Meeting}{\MakeLowercase{\textit{et al.}}: Bare Demo of IEEEtran.cls for Journals}

\maketitle

\begin{abstract}
Different periodical Frequency Selective Surfaces(FSS) whose unit cell are the Split Ring Resonators (SRR)
or related geometries and their corresponding complementary screens were studied. This kind of FSS, sometimes called metasurfaces, have the advantage of avoiding secondary grating lobes because of the small electrical size of the unit cell. The main aim of this paper is to investigate the effects of the metal thickness for these FSSs. It was found that the quality factor decrease for unconnected elements (strips) and increase for connected elements (slots). Besides, as the thickness increases, the resonance frequency decrease for the case of unconnected elements while increases for connected elements. 
\end{abstract}

\IEEEpeerreviewmaketitle

\section{Introduction}
The topic of Frequency Selective Surfaces has a long life and can be considered as classic. For instance, a comprehensive revision of the topic can be found in \cite{Wu}. Roughly speaking, FSSs are filters for electromagnetic waves in free space, which often are considered plane waves. Last years, FSS have received fresh ideas coming from metamaterials. FSS made as periodical array of the Split Ring Resonator (SRR) or related geometries has attracted much attention because of the advantage of a small unit cell compared to the wavelength. It means that secondary grating lobes are avoided. It was demonstrated by Falcone and co-workers \cite{Babinet-Baena, AB-initio} that a periodical 2D array of SRRs behaves as a band stop filter and, according to the Babinet’s principle, its complementary screen behaves as a band pass filter. Both the stopband and the passband are typically narrow, so that the quality factor(Q) is relatively big.

\begin{figure}
  \centering
  \includegraphics[scale=0.3]{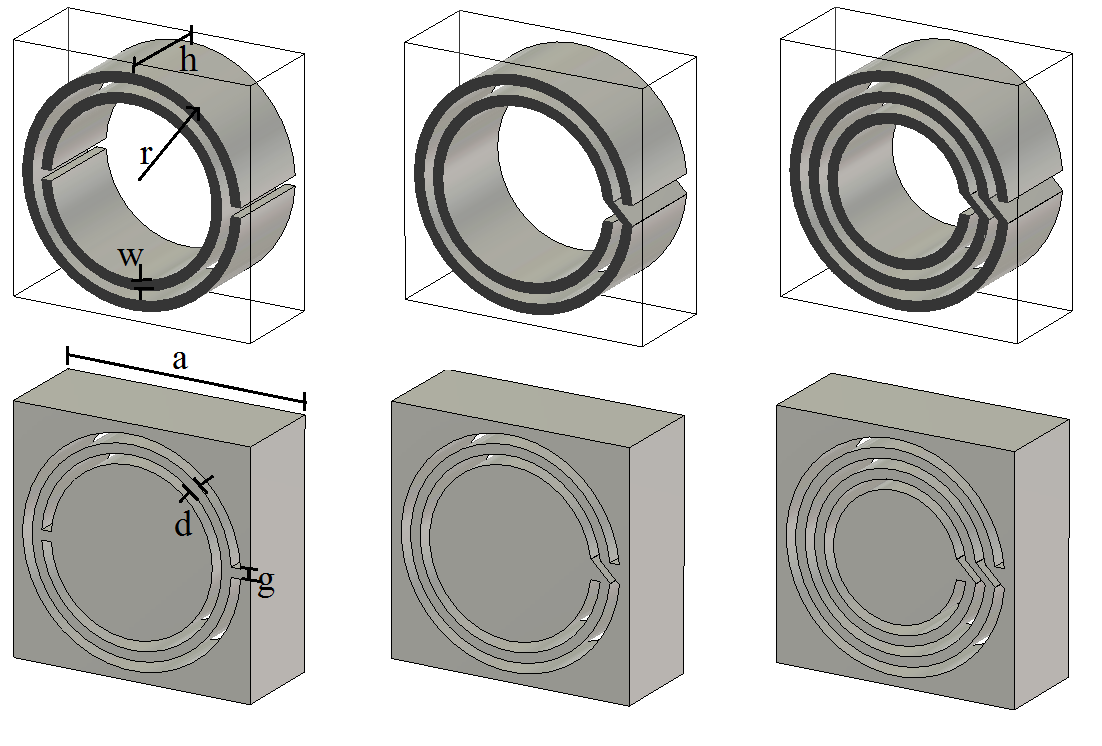} 
   \caption{\label{parametricSRR} Unit cells of the simulated FSSs.First row: the Split Ring Resonator (SRR) and the Spiral Resonators of two and three turns (SR2 and SR3). Second row: complementary unit cell which will be termed as C-SRR, C-SR2, and C-SR3 respectively. The geometric parameters are (all in mm): a = 5,c = d = g =0.2, r = 2.3. The thickness h is varied within the interval 0 to 7 mm.}
\end{figure}

In order to improve the quality factor of the band pass filters, there are two typical routes: cascading several layers or making one single thick layer. For the former case, and for negligible coupling between neighboring layers, the transfer function of a single layer is multiplied by itself as many times as the number of layers. It clearly means an increment of Q for passbands, but a decrement for stopbands. If, instead of the cascade, the structure consist of only one thick layer, then it can be roughly thought as a continuous set of infinite layers and, analogously to the previous case, a thicker FSS will provide higher Q for passbands, but lower for stopbands. However, the Babinet’s principle does not apply for the case of thick metal layers, so that the behavior of the complementary screen cannot be derived directly from the behavior of the original screen. In this work, parametric studies for the thick FSSs shown in \ref{parametricSRR} were carried out. Transmission and reflection coefficients under normal incidence were obtained for many different metal thickness. Special attention was dedicated to understand the effect of the thickness on their Q factors and resonance frequencies.

\section{Preliminary Theoretical Considerations}

For very thin samples and good conductors (ideally infinitesimal thickness and perfect conductors), the basic theory of this metasurfaces is already well explained in \cite{AB-initio}, and references therein. Basically, they firstly analyzed the behavior of the screen made with SRR and after the Babinet’s principle was applied in order to obtain the behavior of the complementary screen. This idea can be easily extended to other related topologies as the Spiral Resonators of 2 and 3 turns (SR2 and SR3). Equivalent circuit models for thin SRR, SR2, SR3, and their complementary counterpart can be found in \cite{SRR-Analysis}. However, as the thickness becomes non negligible, then Babinet’s principle starts to fail. Neither the equivalent circuit models works properly. Thus, we have started to envisaged a theory in order to predict the variations due to the thickness. At the moment of writing this paper, it was to early to show any result of this theories.

\section{Numerical Results}
\begin{figure}
\centering
\includegraphics[scale=1]{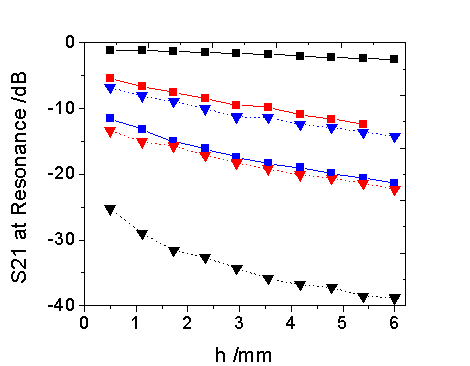}
\caption{\label{tranmission} Transmission coefficient at the resonance frequency vs thickness of the sample.} 
\end{figure}

In order to get some previous insights about the problem, a set of many numerical simulations were run over the commercial software CST Microwave Studio Software. All the FSSs described in Fig. 1 were simulated for different types of metal: Perfect Electric Conductor (PEC), copper, and aluminum. Due
to lack of space, just results for aluminum samples are shown here. Although it is not shown here, narrow stopband were observed for SRR, SR2, and SR3 while narrow passbands for C-SRR, C-SR2, and C-SR3, even for the thickest sample. Figure \ref{tranmission} shows the transmission coefficients at the resonance frequency versus the thickness of the samples. It is worth to note that the best performance is found for the FSSs made of SRR and C-SRR, because they are the nearest to values 0 and 1, respectively. On the other hand, FSSs made of SR2, C-SR2, SR3, and C-SR3 showed very poor performance.

\begin{figure}
                \centering
             \includegraphics[scale=1]{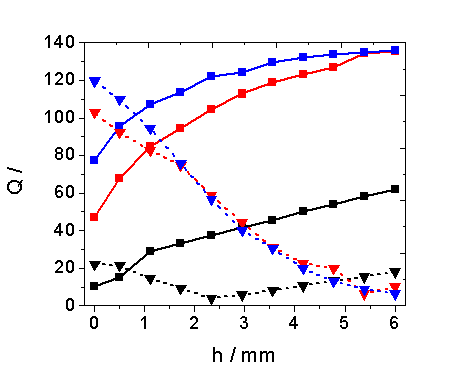} 
\caption{\label{Q-factor} Quality factor (Q) vs thickness of the sample.}
                \label{fig:tiger}
        \end{figure}

\begin{figure}
                \centering
              \includegraphics[scale=1]{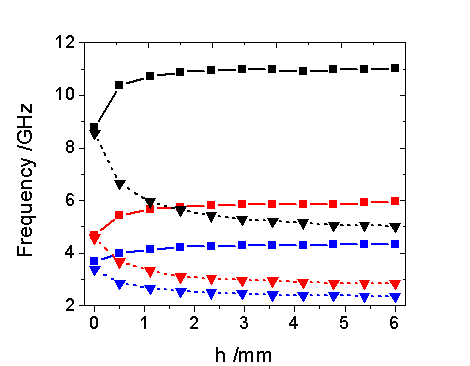} 
\caption{\label{frequency} Resonance frequency vs thickness of the sample h.}
        \end{figure}

Figure \ref{Q-factor} shows the Q factor. As the thickness of the samples are increased, the Q factor drops down for FSSs presenting a stopband (SRR, SR2, and SR3) while it grows for FSSs presenting a passband (C-SRR, C-SR2, and C-SR3). The resonance frequency $f_{0}$ respect to $h$ is depicted in Fig. \ref{frequency}. It is worth to note that, according to the Babinet’s principle, the resonance frequencies for FSSs made of strips and slots elements coincide when the thickness h tends to zero. Interestingly, the resonance frequency of FSSs made of original particles (SRR, SR2, and SR3) shift to lower frequencies while for complementary screens (C-SRR, C-SR2, C-SR3) it shift to higher frequencies.

\section{Conclusions}
The effects of the metal thickness has been studied for the FSSs shown in Fig. \ref{parametricSRR}. It was found that the quality factor and the resonance frequency strongly depends on the thickness. As the thickness is increased, original screens (first row of  Fig. \ref{parametricSRR}) suffer a decrement in both the quality factor and the resonance frequency. Oppositely, complementary screens (second row of Fig. \ref{parametricSRR}) suffer an increment in both magnitudes. Therefore, for further applications, complementary screens might be more promising because of its higher quality factor. For instance, applications such as passband radomes to hide directive antennas need very high Q factor which could be obtained by using thick complementary screens. It makes this study interesting from a practical standpoint.

\end{document}